\documentclass[10pt]{iopart}

\usepackage{iopams}  
\usepackage{graphicx}
\usepackage[margin=10pt,font=small,labelfont=bf,labelsep=endash]{caption}

\begin{document}

\title[]{Effect of scintillator geometry on the energy resolution and efficiency of MAST neutron camera detectors}

\author{M. Cecconello$^{1,2}$}

\address{$^1$Department of Physics and Astronomy, Uppsala University, SE-751 05 Uppsala, Sweden}
\address{$^2$Department of Physics, Durham University, Durham, DH1 3LE, UK}
\ead{marco.cecconello@physics.uu.se}

\begin{abstract}
The efficiency and energy resolution of the MAST neutron camera detectors, based on liquid scintillator of the EJ-301 type, is lower than cylindrical detector of similar size. An experimental investigation has identified light attenuation as the main cause of this deterioration. This finding is supported by Monte Carlo electron and photon transport calculations. 
\end{abstract}

%

%
%
%
\ioptwocol

\section{Introduction}
\label{sec:Intro}
In thermonuclear fusion devices, liquid scintillators are commonly used for detection of neutron emission from deuterium-deuterium and deuterium-tritium nuclear reactions. They are usually employed as simple counters in collimated geometry where collimation is achieved with a thick shielding of low Z elements (typically concrete or polyethylene) having the double purpose to provide a direct view of well defined regions of neutron emission from within the plasma volume and to suppressed scattered neutrons. For example, on the Mega Ampere Spherical Tokamak (MAST) \cite{Darke1995} an array of four collimated neutron flux monitors based on the EJ-301 liquid scintillator \cite{Cecconello2014} were used to correlate the temporal evolution of the spatial distribution of fusion reaction rates to instabilities in the fusion plasma itself \cite{Cecconello_2014}. 
The response function of liquid scintillators to neutrons, discussed in detail in \cite{Cecconello2019}, is dominated by the elastic scattering with the H nucleus, the non-linear conversion of the recoil proton energy into scintillation light, the role of multi-scattering interactions which depends on the thickness of the active volume and the scintillation light collection and transport to the photocathode for its conversion into a voltage signal. 
Depending on the specific design of the detector, and in particular the distance between the photocathode(s) and the point where the scintillation light is emitted, light attenuation through the scintillation medium can affect the amplitude of the photomultiplier output voltage signal for the same energy being deposited in the medium.
This reduction in amplitude results in a shift of the pulse height spectrum (PHS) to lower amplitudes and, in the case of detection system with a fixed acquisition threshold, in the reduction of the absolute detection efficiency. This effect was studied in the detail with collimated $ \gamma $-rays sources and it was found to be significant even for relatively small detectors \cite{KUIJPER196656}, \cite{DELEO1974559} and \cite{ANNAND1983421}.
Knowledge of the absolute detection efficiency is fundamental for the correct measurement of the power produced by fusion reactions and ultimately of the performance of fusion devices. For example, the difference between measured and simulated neutron rates at JET and MAST has been the focus of detailed studies leading to questioning both the modeling of the neutron emission and the accuracy of the absolute calibration of neutron diagnostics \cite{Weisen_2017} and \cite{Cecconello_2018}.  
Light attenuation leads to an additional complication, namely the degradation of the energy resolution \cite{SCHOLERMANN198025} specifically for rectangular geometries \cite{Smith_1972}. In particular this effect is worsened when the liquid scintillator is exposed to an extended source of radiation impinging uniformly on the whole active volume. This is the typical case for the neutron detectors installed at MAST. In this situation, even assuming a flux of mono-energetic neutrons impinging on the detectors, the observed PHS is the sum of the PHSs generated by the scintillation light emitted at different positions along the scintillator each multiplied by an attenuation coefficient that depends of the distance from the photomultiplier.
The overall effect is to reduce the energy resolution of the PHS by broadening of the edge corresponding to the energy deposited by recoil protons generated in ``head on'' collisions in the case of incident neutrons or to the recoil electron Compton edge in case of incident $ \gamma $-rays. 
As a result, the limited capabilities of liquid scintillators as neutron energy spectrometers in determining key parameters as the fuel deuterium and tritium ions temperature, the presence of fast ions with energies above the thermal plasma and the the plasma rotation are further deteriorated.  

The aim of this study is to estimate the reduction in the detection efficiency of the detectors used in the collimated neutron flux monitor installed at MAST and to demonstrate how their poor energy resolution, compared with other EJ-301 liquid scintillators of comparable dimensions but different geometry, is directly correlated to the attenuation of the scintillation light. The study has been carried out with a $^{22}$Na $ \gamma $-rays source rather than mono-energetic neutron sources as the latter are much less available without any loss of validity. 
Section \ref{sec:SEK} is dedicated to a brief description of the liquid scintillator detectors on MAST and of their response function to $ \gamma $-rays as well as of the experimental set-up used to study the effect of light attenuation while the findings are presented in section \ref{sec:results}. The obtained results are interpreted in section \ref{sec:MCNP} using the General Monte Carlo N-Particle (MCNP) transport code \cite{MCNP6}. Discussions and conclusions are presented in section \ref{sec:DC}.

\section{Liquid scintillator properties and collimated setup}
\label{sec:SEK}
\begin{figure}[!htp]
	\begin{center}
		\includegraphics[scale=0.5]{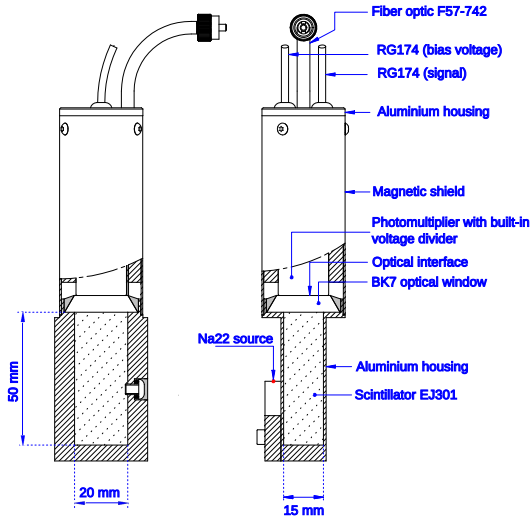}
	\end{center}
	\caption{Front and side section views of the detector: the liquid scintillator EJ301 has an active volume of $50 \times 20 \times 15$ mm$^3$. Light transmission to the photomultiplier is via a BK7 window. The red dot indicates the location of the $^{22}$Na $ \gamma $-rays source used for calibration.}
	\label{fig:detector}
\end{figure}

\begin{figure}[!htp]
	\begin{center}
		\includegraphics[scale=0.4]{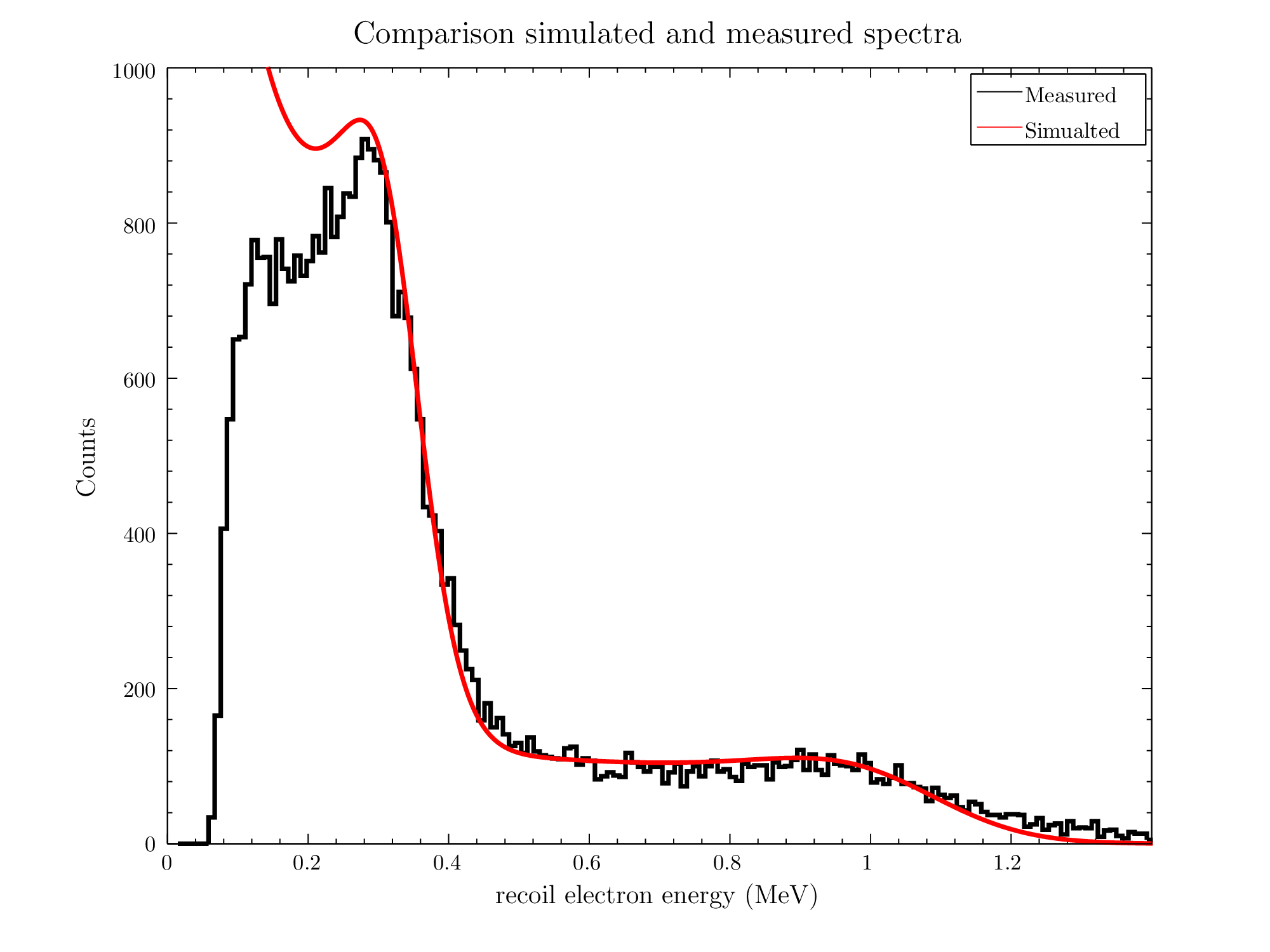}
	\end{center}
	\caption{Comparison between the measured recoil electron PHS with the collimated narrow beam $^{22}$Na $\gamma$-rays source and the \texttt{MCNP} predicted spectrum when fitting the PHS spectrum itself.}
	\label{figure:SEKMCNPfit}
\end{figure}

The detectors used in the collimated neutron flux monitor installed at MAST are based on the EJ-301 liquid scintillator (equivalent to the more know NE213) with an active volume of $50 \times 20 \times 15$ mm$^3$ inside an aluminum housing. The inside walls of the aluminum chamber containing the liquid scintillator are coated with a TiO$ _2 $ reflective coating.  The scintillation light is collected via a BK7 optical window mounted on the short side and converted into a voltage signal via the Hamamatsu R5611A photomultiplier. A schematic of the detector is shown in figure \ref{fig:detector}.
The active volume is exposed to a collimated neutron flux of cross-sectional area of $50 \times 20$ mm$^2$ through the 1 mm thick aluminum housing. The choice of this unusual shape over the more traditional cylindrical one for the scintillator was motivated by the requirement of hight count rates and high spatial resolution. In MAST plasmas, the neutron source is elongated in the vertical direction with a corresponding larger change in the neutron emissivity profile in the horizontal direction.
On the back surface, a $^{22}$Na $ \gamma $-rays source can be mounted: the location of the point-like source is indicated by the red dot in figure \ref{fig:detector}. This radioactive source is used to carry out the energy calibration of the detector linking the output voltage to the recoil electron Compton edge energy and, via the relation between the electron and proton light output functions, to the energy deposited in the detector by the neutrons.
From this figure, it is clear that the $ \gamma $-rays source is not collimated, i.e. the whole active volume is exposed to the $ \gamma $-rays (and neutrons). This is the configuration in which the detectors are operated on MAST. An example of the resulting recoil electron PHS in this configuration is shown in figure \ref{figure:SEKMCNPfit}. 

The response of the liquid scintillator to $\gamma$-rays is calculated with MCNP. In this model, the detector is in empty space and the $\gamma$-ray source is approximated as an isotropic point source located as shown figure \ref{fig:detector} emitting $ \gamma $-rays with 0.511 and 1.275 MeV energy. 
The pulse height distribution is calculated using the F8 tally card to determine the electron energy deposition in the energy range 0 - 2.5 MeV in the liquid scintillator only. Photons are transported in all the detector's parts while electron transport is only followed in the liquid scintillator and suppressed elsewhere. The transport of the scintillation light, and therefore its attenuation, is not modeled by MCNP.
The calculated PHS is convoluted with an energy dependent resolution function $ \Delta E/E\ = \Delta L/L = \sqrt{\alpha^2 + \beta^2/L + \gamma^2/L^2 } $ \cite{DIETZE} for a comparison with the measured one: the results is shown in figure \ref{figure:SEKMCNPfit} by the red curve with $ \alpha = 21.7 $, $ \beta = 8.4 $ and $ \gamma = 6.8 $.
The energy resolution function parameters $ \alpha$, $ \beta $ and $ \gamma $ above have been determined via least square fit of the experimentally measured energy resolution at the position of two Compton edges (similar values are obtained by forcing $ \gamma =0 $ ). The resulting energy resolution is about 2-3 times higher than the energy resolution observed for cylindrical scintillators of comparable volume \cite{DIETZE}. 

\begin{figure}[!htp]
	\begin{center}
		\includegraphics[scale=0.4]{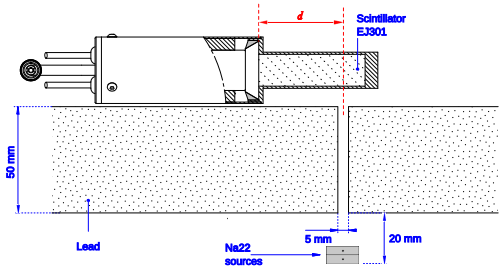}
	\end{center}
	\caption{Experimental setup for the measurement of recoil electron pulse height spectra with collimated $^{22}$Na $ \gamma $-rays source using 5 cm thick lead bricks with a 0.5 diameter collimator. By moving the detector, the $ \gamma $-rays beam illuminates the active volume at different distances (indicate by $ d $) from the BK7 optical window.   }
	\label{fig:collimation_setup}
\end{figure}

In order to irradiate the liquid scintillator with a collimated beam of $ \gamma $-rays the experimental set-up shown in figure \ref{fig:collimation_setup} was used. The calibration source was removed and instead two $^{22}$Na $\gamma$-rays sources were stacked on top of each other to increase the count rate (total activity of 33.7 kBq). The detector was shielded by lead bricks 50 mm thick extending well beyond its lateral dimensions. A hole of 5 m diameter drilled in the lead was used to collimate the $ \gamma $-rays. 
The detector was then moved along its axis and parallel to the lead and recoil electron PHS spectra were collected as a function of the distance $ d $ from the BK7 window: in this way, the collimated $ \gamma $-rays were incident on a small area centered on the axis of the detector.
A reference PHS was also acquired with the two $^{22}$Na sources in place by replacing the lead brick with the collimator with a solid one of the same dimensions: this PHS was then subtracted to all the measured PHS.  
The photomultiplier was operated at a constant negative voltage of -950 V and the pulses from the detectors were recoded with a 250 MSamples/s diqitial acquisition board. 
The individual pulses were then processed for baseline correction, pulse shape discrimination and the calculation of the total charge in each pulse which is proportional to the intensity of the scintillation light and therefore to the energy deposited by the $ \gamma $-rays.

\section{Results}
\label{sec:results}
\begin{figure}[!htp]
	\begin{center}
		\includegraphics[scale=0.4]{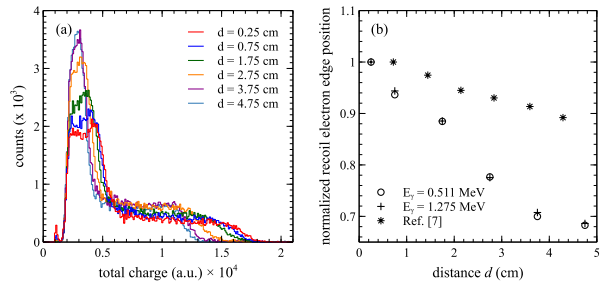}
	\end{center}
	\caption{Panel (a): pulse height spectra of the total charge of the recoil electron pulses when the detector is illuminated by a collimated beam of $^{22}$Na $\gamma  $-rays as function of the distance $ d $ from the BK7 optical window. The effect of light attenuation is clearly visible. Panel (b): position of the recoil electron Compton edges (open circle for $ E_\gamma = 0.511 $ keV and cross for $ E_\gamma = 1.275$ MeV) normalized to the corresponding values at $ d = 0.25$ cm as a function of the distance $ d $ compared with data from \cite{ANNAND1983421}.}
	\label{fig:collimated_PHS_Attenuation}
\end{figure}

Electron recoil pulses were measured for $ d = 0.25, 0.75, 1.75, 2.75, 3.75, 4.75 $ cm from the BK7 window. The resulting six PHS, based on the total charge (voltage pulse integrated over the pulse duration), are shown in panel (a) of figure \ref{fig:collimated_PHS_Attenuation}. As it can be seen, the further away the PMT is from the collimated $ \gamma $-rays beam, the smaller the collected total charge is, a clear indication of the effect of the attenuation of the scintillation light in the scintillator itself.
Panel (b) of figure \ref{fig:collimated_PHS_Attenuation} shows how the normalized position of the recoil electron Compton edge depends on the distance $ d $. The Compton edge for each irradiation position $ E_C(d_i) $ is normalized to the its value for $ d = 0.25 $ cm: this was motivated by the assumption that light attenuation for this irradiation position is negligible.
The position of the Compton edge is determined as the total charge corresponding to half the counts in the Compton edge maximum.
As it can be seen, the scintillation light for both Compton edges is attenuated by the same amount, with an attenuation factor of approximately 0.3 cm$^{-1}$ which is much larger than what is observed for cylindrical detectors: for example, the normalized recoil edge position for $^{137}$Cs $ \gamma $-rays source using a NE213 scintillator of 5 cm height and diameter surrounded by TiO$_2$ paint has an attenuation of $\approx$ 0.1 cm$^{-1}$ as shown in panel (b) of figure \ref{fig:collimated_PHS_Attenuation}.
This large difference can be partially explained by the different geometry which result in a larger number of reflections in the case of the detector here studied. A simple estimate of this effect can be obtained assuming a detector of height $ h $ and lateral dimension $ a $. 
Assuming that light is emitted at the opposite surface of the light collection optics at an angle $ \theta $  with respect to the normal to that surface and in the direction of the PMT, then the number of reflections it undergoes before reaching the PMT is $ n \approx (h/a) \tan(\theta) $ and the intensity of the light after $ n $ reflections is $ I_n = r^n I_0 $ where $ r $ is the average reflectivity of the coating and $ I_0 $ the initial intensity.
For a cylinder with $ h = a $ as in the case of figure \ref{fig:collimated_PHS_Attenuation}, the ratio $R = I_n /  I_0$ is equal to $ R_1 = r^{\tan(\theta)} $ while for the scintillator studied here, where $ a = 3h/10 $, it is $ R_2 = R_1^{10/3}$. Assuming $ r \approx 0.96 $ at 425 nm \cite{ANNAND1983421} and $ \theta = \pi/4 $ then $ R_1 = 0.96 $ and $ R_2 = 0.87 $. Note that the total length traveled by the light depends on the scintillator height $ h $ and angle $ \theta $ but not on the lateral dimension $ a $ and is given by $ h/\cos(\theta) $. The light attenuation is then the same in both cases.
\begin{figure}[!htp]
	\begin{center}
		\includegraphics[scale=0.4]{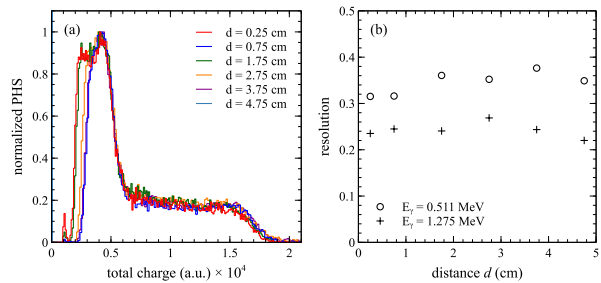}
	\end{center}
	\caption{Panel(a): the same as in panel (a) but with the PHS normalized to the respective maximum value and with the total charged corrected for the attenuation (see panel (a) of figure \ref{fig:collimated_PHS_Attenuation}). Panel(b): resolution at the Compton edge energy for  $ E_\gamma = 0.511, 1.275$ MeV as function of the distance $ d $.}
	\label{fig:collimated_PHS_Resolution}
\end{figure}

Panel (a) of figure \ref{fig:collimated_PHS_Resolution} shows the same PHS shown in panel (a) of figure \ref{fig:collimated_PHS_Attenuation} but with the total charge corrected for the attenuation shown in panel (b) of the same figure clearing indicating that attenuation due to light absorption in the liquid scintillator and to partial reflection at the wall of the housing are responsible for the modification of the shape of the PHS. 
Note also that the Compton edges are much sharper compared to those shown in figure \ref{figure:SEKMCNPfit}. 
In this case, such broad edges are now well understood since the PHS obtained with the uncollimated $ \gamma $-rays source can be thought off as the sum of the PHS obtained with a collimated source at different location along the scintillator. 
The different locations of the Compton edges due to the attenuation cause the broadening of the PHS in the case of the uncollimated source. This is further confirmed in section \ref{sec:MCNP}.
The detector resolution, calculated as the ration between the total charge at 10 and 90 \% of the Compton edge maximum divided by the total charge at 50 \% of this value, however does not depend on the distance $ d $ as shown in panel (b) of figure \ref{fig:collimated_PHS_Resolution} and it is still much larger of similar sized cylindrical collimators.

\section{MCNP modelling}
\label{sec:MCNP}
\begin{figure}[!htp]
	\begin{center}
		\includegraphics[scale=0.4]{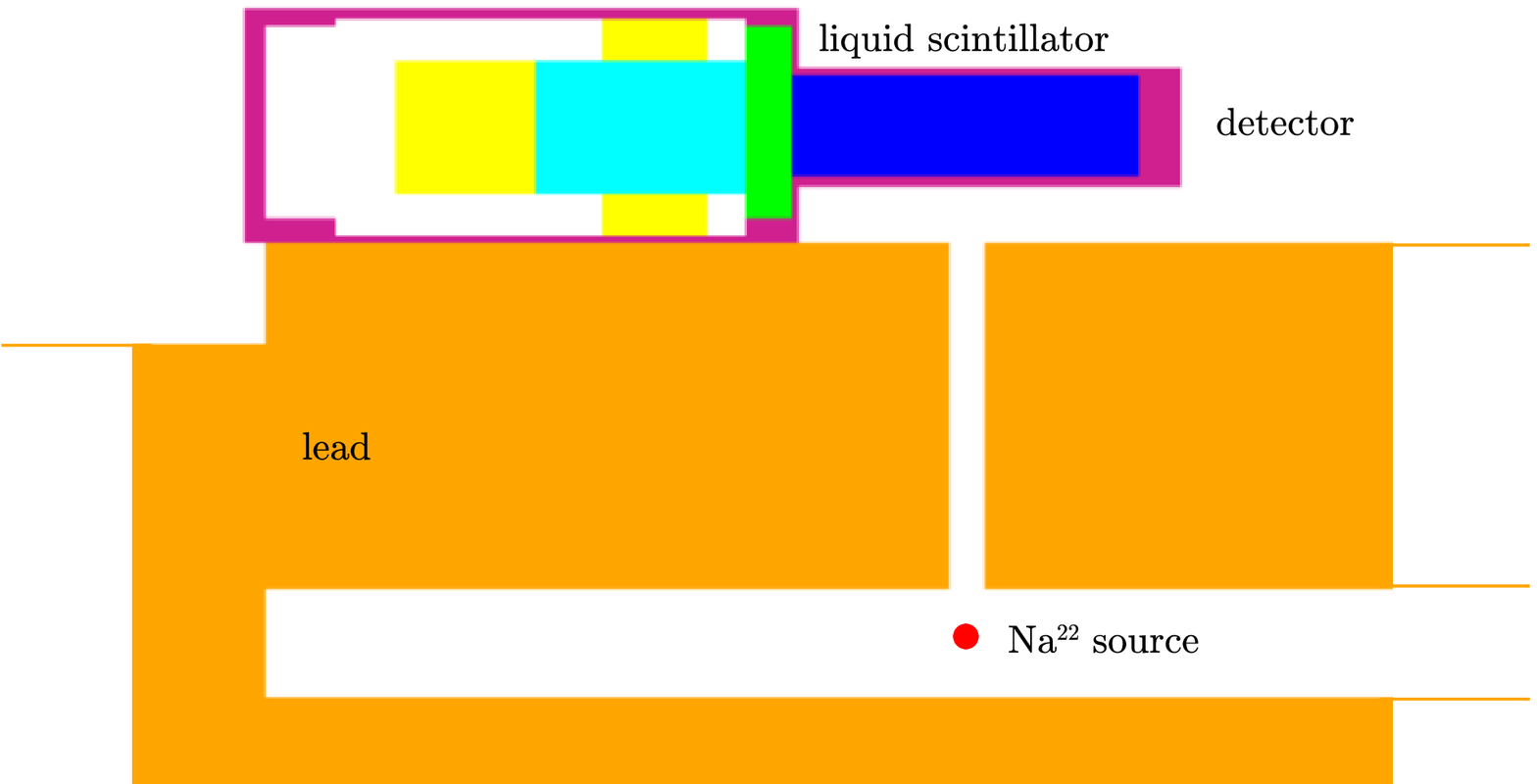}
	\end{center}
	\caption{Comparison between the measured recoil electron PHS with the collimated narrow beam $^{22}$Na $\gamma$-rays source and the \texttt{MCNP} predicted spectrum when fitting the PHS spectrum itself.}
	\label{figure:MCNP_Collimated}
\end{figure}

Figure \ref{figure:MCNP_Collimated} shows the MCNP model of the liquid scintillator for $ d = 2.5 $ cm used for the measurement of the PHS with a collimated $\gamma$-rays source.
In this simulation, the transport of $ \gamma $-rays is followed everywhere within the model while electron transport is tracked in the liquid scintillator, the housin and the BK7 optical waveguide.

\begin{figure}[!htp]
	\begin{center}
		\includegraphics[scale=0.4]{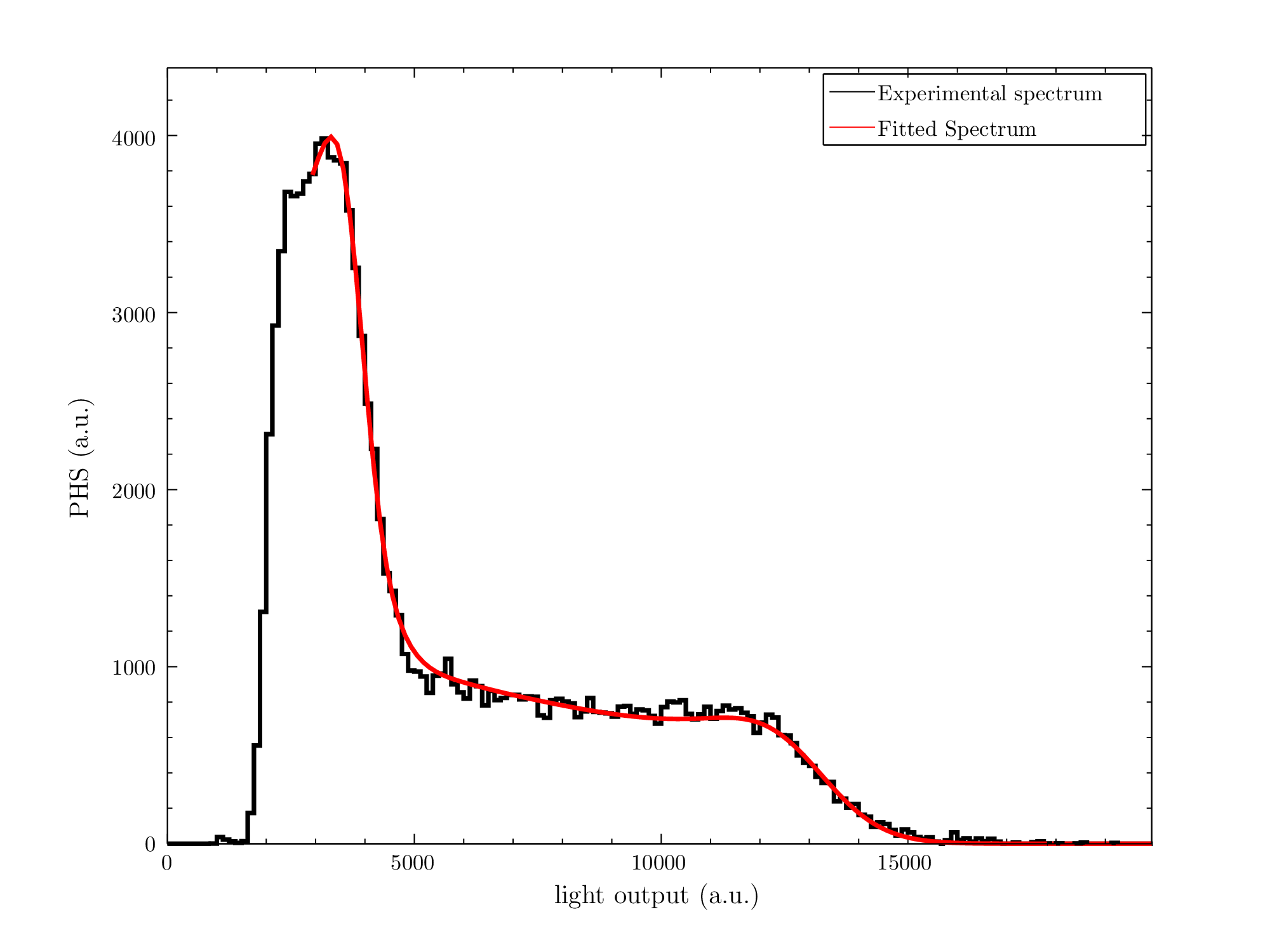}
	\end{center}
	\caption{Comparison between the measured recoil electron PHS with the collimated narrow beam $^{22}$Na $\gamma$-rays source and the \texttt{MCNP} predicted spectrum when fitting the PHS spectrum itself.}
	\label{figure:MCNP_collimated_PHS_Attenuation}
\end{figure}

The resulting PHS is then convoluted with the energy resolution function whose parameters $ \alpha $, $ \beta $ and $ \gamma $ are obtained by a least square fitting to the measured PHS. The comparison between the simulated and measured PHS is shown in figure \ref{figure:MCNP_collimated_PHS_Attenuation}. As it can be seen, in this case the agreement between the two is improved compared to the result shown in figure \ref{figure:SEKMCNPfit}, highlighting the importance of proper modelling of the transport of the scintillation light \cite{TAJIK2013104}.

\begin{figure}[!htp]
	\begin{center}
		\includegraphics[scale=0.5]{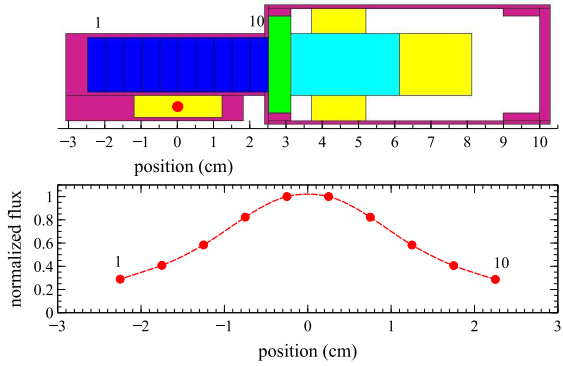}
	\end{center}
	\caption{Top panel: MCNP model of the detector with the liquid scintillator indicated in blue; the red dot represent the location of the $^{22}$Na source inside its plastic holder (yellow region). Bottom panel:  $ \gamma $-rays flux in different regions, numbered 1 to 10, of the scintillator.}
	\label{fig:MCNP_Fluxes}
\end{figure}

\begin{figure}[!htp]
	\begin{center}
		\includegraphics[scale=0.5]{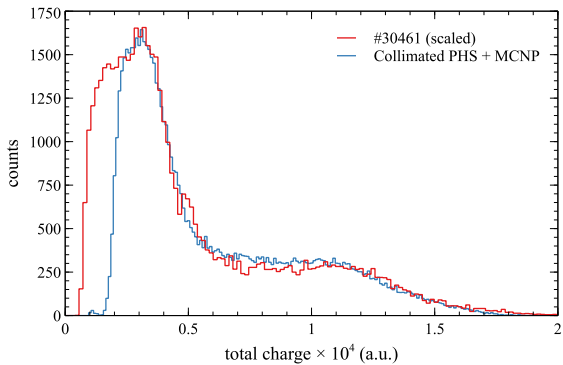}
	\end{center}
	\caption{The recoil electron PHS for pulse \#30461 with the $^{22}$Na $\gamma$-rays source located in its reference position (in red) compared with the average of the collimated PHS weighted by the normalized flux calculated with MCNP (in blue).}
	\label{fig:comparison}
\end{figure}

As a final verification of the interpretation of the broadened PHS for uncollimated sources provided in section \ref{sec:results}, the uncollimated geometry has been modeled in MCNP as shown in figure \ref{fig:MCNP_Fluxes}.
The liquid scintillator region has been divided in 10 cells and the photon flux has been calculated in each of the cells.
The $ \gamma $-rays flux as a function of the position along the scintillator is shown in the bottom panel of figure \ref{fig:MCNP_Fluxes} and are used to weight the experimental PHS obtained with the collimated source. 
The weighted average of these six PHS agrees quite well with the uncollimated PHS as shown in figure \ref{fig:comparison}.


\section{Discussion}
\label{sec:DC}
The results presented in the previous section clearly show how the geometry of the detectors installed on the neutron camera on MAST affected negatively their energy resolution thus limiting the possibility of using them to carry out more sophisticated studies of the plasma ion temperature, the plasma rotation and the changes in the fast ion distribution induced by plasma instabilities.
In addition, such detectors are affected by a reduction in the detection efficiency due to the shift towards lower amplitudes of the light pulses generated far from the PMT and to a constant acquisition threshold.
This effect can be clearly seen in panel (a) of figure \ref{fig:collimated_PHS_Attenuation}: the counts in the lower total charge part of the PHS increases with the distance as more and more attenuated scintillation light pulses reach the PMT. 
A simple model describes the reduction in the efficiency as $ \epsilon(a) = (aQ - Q_T)/a $ where $ a $ is the relative reduction in the scintillation light intensity as a function of the distance from the PMT and $ Q_T $ is the total charge corresponding to the acquisition threshold. 
For the detector here studied, the relative reduction in efficiency for $ d = 5 $ cm is about 4 \% compared to the efficiency for $a = 1$ ($ d = 0 $ cm) resulting in an average reduction in efficiency of approximately 2 \%. This reduction in the efficiency is clearly insufficient to contribute to the discrepancy between measured and simulated neutron rates on MAST \cite{Cecconello_2018} which has been partially accounted for in a recent study \cite{Sperduti_2020} and linked to the modeling of the fast ion dynamics in the plasma. 
Although the reduction in the efficiency is negligible, due to their poor resolution the neutron camera upgrade is equipped with more traditional cylindrical liquid scintillator with a height of 1.5 cm and a diameter of 3 cm which have a resolution of about  8 \% at the Compton edge for $ \gamma $-rays of 1.275 MeV energy \cite{NCU}.

\ack
The author wish to thanks S. Conroy and H. Hjalmarsson for fruitful discussions. 


\end{document}